# The search for a strategy for mankind to survive the solar Red Giant catastrophe

M. Taube* and W. Seifritz**

Abstract

In around 5 gigayears our Sun will grow to a Red Giant and will swallow Earth. In this article we present a rescue-plan for the whole mankind, including its inherited habitat Earth. The plan is subdivided into two parts:

If we content ourselves to survive only the next 5 Gy (until the shell-burning phase of our Sun) we propose to construct some kind of parasol to shadow Earth rather than to move Earth to an outer orbit using asteroids for the swing-by technique. The position of the parasol will be the (inner) Lagrange Point L1 or depending on the radiation pressure onto the parasol, in a levitated orbit a bit nearer to the Sun. The sidereal period of these orbits is one year guaranteeing a permanent eclipse of Earth.

If we want to survive also the time beyond the next 5 Gy, where Sun's luminosity and radius increase hundred fold and oscillate until our Sun develops finally into a White Dwarf, we have to shift Earth into the Kuiper Belt (50 AU) by means of the swing-by technique whereby the overwhelming part of the gravitational energy needed will be transferred from the orbit of Jupiter and Saturn. During this journey of about 10 megayears or more Earth must be illuminated by an artificial light source because of the intensity by our sunlight becomes the smaller the distance increases. A ring of DD-fusion power stations outstretched on Moon's orbit should produce the necessary 175 PW of visible light with which Earth has to be illuminated.

In the Kuiper Belt Earth will be brought into an orbit of an artificial Sun, an ArtSun formed in the meantime by the fusion of gaseous Jupiter-like planets imported from other planetary systems in the neighborhood. Depending on the total hydrogen mass we will bring together a G5-class star (very similar to our Sun) or a M-class star (Red Dwarf similar to the Gliese 581) with a mass of one third of that of our Sun can be formed - the latter one with the advantage that its life-time is about an order of magnitude higher than that of our Sun.

---------------------------------------------------

* Bollackerweg 10, CH-8956 Killwangen, Switzerland, taubemiec@netwings.ch
** Mülacherstr 44, CH-5212 Hausen, Switzerland

## Introduction

We base our reflections on the new publications about the distant future of our Sun (Schröder, 2008) presented in Table 1 from its early phase (age: 0.00 Gy) until its Red Giant stage (age: 12.3 Gy). The formulation of out question is: Can mankind as a whole survive this Red Giant evolution or at least some of the first phases and what would be the appropriate strategy?

**Table 1. Main characteristic of Sun in the distant future**
(according to Schröder, 2008)

| Phases | Age (Gy) | Luminosity $L/L_o$ | Surface temperature, $T_{eff,}$ [K] | Radius $R/R_o$ | Mass $M/M_o$ | Comments |
|---|---|---|---|---|---|---|
| Early | 0.00 | 0.7 | 5 596 | 0.89 | 1.000 | Central H-burning |
| Present | 4.58 | 1.00 | 5 774 | 1.00 | 1.00 | |
| Later phases | 7.13 | 1.26 | 5 820 | 1.11 | 1.00 | |
| | 10.00 | 1.84 | 5 751 | 1.37 | 1.000 | |
| Red Giant phase | 12.17 | 2 730 | 2 602 | 256 | 0.668 | H-shell burning. Expansion of the outer layer, start of mass- loss, high luminosity of the 'first giant branch', |
| | 12.17 | 53.7 | 4 667 | 11.2 | 0.668 | |
| | 12.3 | 2 090 | 3 200 | 149 | 0.546 | |
| | 12.3 | 4 170 | 3 467 | 179 | 0.544 | low temperature, He-burning |

$L_o$: Solar luminosity, at present, Note : 1.00 AU (Astronomical Unit) = 215 $R_o$

Historically one of the first chains of reasoning was to build a giant space ship for a limited number of survivors, called interstellar colonists, to escape our Solar System. For example Dyson (Dyson, 1968) proposed an 'Ablation Space Ship' for a payload of 10 000 tons with a mission velocity of 10 000 km/s. He showed that hydrogen bomb detonations (about 300 000 bombs) could take over chemical propulsion as an energy source for long-range space travel. The bombs would explode behind a huge hollow copper hemisphere covered with some ablating substance protecting the underlying structure and a small layer of the ablating material would be vaporized by each burst - thus increasing the momentum transfer. Although it might be technically possible to design such a space-ship in the next centuries it has some inherent disadvantages such as:

a) the relatively small number of people which could be evacuated (compared with the 6.7 billion today),

b) the giving-up the photosynthesis supported habitat 'Earth' and

c) the fact that we do not really know about the new places to which the colonists should go finally. We reject this idea not only because of the inherent uncertainty of the place of destination but mainly because of the leaving in the lurch of the majority of mankind and of its whole biosphere.

A seemingly better idea emerged during the last decades: Korycansky (Korycansky, 2001) proposed the lift up Earth's orbit (1 AU) to the orbital radius of Mars (~1.52 AU) or something beyond by using gravitational assists to transfer orbital energy from Jupiter (and perhaps Saturn) to Earth using repeatedly highly elliptical orbit (~300 AU ) asteroids from the Kuiper Belt by 'swing-by' manoeuvres. An object with a mass of $10^{22}$ g (diameter 100-150 km) every 6000 years (on average) is guided in such manner that it passes close to the leading limb of Earth, i.e. at the trailing side of its orbital path (distance ~10 000 km). It could do the above job within about 1 Gy if the resulting out-bound trajectory crosses afterwards the orbit of Jupiter to pick up energy that was lost to Earth.

The theory of the 'swing-by'-technique is well known today and used to accelerate satellites out of the solar system. It is simply based on the conservation of energy and the equation of momentum transfer of such a three-body problem in the heliocentric system whereby the Sun possesses zero velocity. It will not be repeated here; but to get an idea of the huge amounts of energy involved when shifting the orbit of a planet into higher orbit, we make the following estimation: Based on the virial theorem in a conservative system with a 1/r-potential the sum of kinetic and potential energy, i.e. the Hamiltonian, of a planet with mass m is given by

$$E = -\frac{\gamma.\mathrm{m}.\mathrm{M}}{2a} \text{ [erg]} \tag{1}$$

where $\gamma = 6.67 \cdot 10^{-8}$ cm³/(g.s²) = gravitational constant, $M=1.99 \cdot 10^{33}$ g = mass of Sun, $m=5.95.10^{27}$ g = mass of Earth and $a = 149.6 \cdot 10^{11}$ cm = 1 AU is the mean radius of its orbit, when we measure the RHS of eq (1) in the CGS-system. The negative sign stems from the fact that we have to do with a bound state. Therefore, if we shift Earth from to an higher orbit a' (a'>a) we need an amount of energy, $\Delta E$, given by

$$\Delta E = \frac{\gamma.\mathrm{m}.\mathrm{M}}{2} \left(\frac{1}{a} - \frac{1}{a'}\right) \text{ [erg]} \tag{2}$$

With $a'=227.9 \cdot 10^{11}$ cm = mean orbital radius of Mars, for instance, we obtain $\Delta E = 9.07 \cdot 10^{39}$ erg = $2.87 \cdot 10^{10}$ PWy.

To move Earth to the Kuiper Belt (a’>>a, $a' \rightarrow \infty$) we obtain $\Delta E = 8.4 \cdot 10^{10}$ PWy.

In a *Gedankenexperiment* we could imagine to install some kind of a machine thrusting Earth into Mars’s orbit and we would allow a time span of one Gy we obtain a mean power of P=28.7 PW which corresponds to about 3000 times the present anthropogenic energy production of mankind - an unimaginable undertaking: Even if we could use nuclear fuels our atmosphere would be destroyed in a short time by such a continuous Hohmann-Transfer.

This estimation shows very clearly that the ‘swing by’- method (Korycansky, 2001) , shuffling gravitational energy from other planets orbits to Earth, is the more suitable technique for a move of Earth - although we have to keep in mind that we need a certain amount of energy to guide the asteroid correspondingly. But this amount of energy will be very small compared with the above-mentioned orbital energies $\Delta E$. We acknowledge, therefore, the progress made by the ‘swing-by’ technique to move Earth outwards at least for the next 5 Gy in such a way that the solar constant remains essentially constant. The solar surface temperature $T_{eff}$ as well as the solar mass remain practically constant in this time span, the solar radius will be only about one third larger but the luminosity increases by a factor of 1.82 (see Table 1). The task will be to increase the orbit of Earth by a factor of $\sqrt{L/L_o} = \sqrt{1.87} = 1.37$ within the next ~5 Gy to maintain the present influx of solar energy, i.e. the present solar constant. Indeed this rescue plan could work to survive the next 5 Gy. But what then, when solar luminosity increases over thousand fold and the solar radius exceeds the orbit of Earth?

According to Schröder (Table 1) in the following 2 Gy Sun’s luminosity increases ‘suddenly’ by a factor of 2730, the size of Sun blows-up by a factor of 256 and its mass reduces by one third because the H-shell burning phase begins. We had not only to move Earth’s orbit by a of factor of $\sqrt{2730} =$ ~52, i.e. far beyond Pluto’s orbit (~40 AU) that is the Kuiper Belt, but the solar wind (hot protons) resulting from Sun’s mass loss would destroy our habitat. The lifting-up of the orbit of Earth to that of the non-engulfed Mars does not help so much because in all the discussions so far one has forgotten to take into account the destructive power of the resulting strong solar wind.

Furthermore, there will be another problem. Since the solar surface temperature $T_{eff}$ reduces in this time by more than a factor 2 it will be questionable if we could maintain the photosynthesis process, working presently with two photons to split water. If the solar surface temperature goes down to about 2600 K we will lose the hard part of spectrum and it will be questionable if photosynthesis can work as efficiently as today. Perhaps genetic engineering will allow us to introduce a three-photon photosynthesis process. Our understanding of life on a planet is not only that we can dispose on liquid water but also the availability of a suitable solar spectrum to maintain photosynthesis (Taube, 1965, 1968).

Already 25 years ago, one of us (Taube, 1984) (see also Cathcart, 1983, Fogg, 1989, Prantzos, 2000) , proposed the shifting of Earth's to an orbit near Jupiter's orbit, significantly before Sun achieves the phase of the Red Giant. During this shift a 'synthetic solar light source' is required, with a power of around 175 TW directed towards Earth and fueled with deuterium extracted from Jupiter. The Earth was proposed to be shifted by means of 'recoil rockets' constructed round the terrestrial equator, but above the terrestrial atmosphere. In those days the concept of the 'swing-by' was not yet enough tested.

Our conclusion is, therefore, the following: The approach to move Earth by successive 'swing-by' manoeuvres of asteroids to survive the next 5 Gy may be technically possible – but not more. The story of life on Earth would end definitively after 5 Gy and survival beyond this time scale will be impossible. Although the 'swing-by'- idea is an elegant method it will have some inherent problems which must be mentioned, for instance, such as the problem of a collision of the 'swing-by' asteroid with Earth, the danger of a spring tide of the oceans during the encounter due to the sphere of gravity of the nearby asteroids etc. One has to note that the kinetic energy of an object with $10^{22}$ g, passing nearby Earth with velocity of the realm of 50 km/s, is about $1.25.10^{35}$ erg corresponding to about 3000 Petatons TNT-equivalent. This means that an accidental impact on Earth would be its end. Furthermore, the attractive force of this object passing Earth in a distance of about 10 000 km onto the oceans-water mass would be about the twice as high as the influence of the Moon onto these water masses – thus increasing the probability of a spring-tide. All these problems have to be analyzed in detail when the 'swing-by' method will be really applied.

Due this general survey we propose in the following two rescue-plans:

a) a survival plan for 'only' the next 5 Gy, based on a less ambitious approach than the 'swing-by' method, namely through shadowing Earth by a parasol in the inner Lagrange Point $L_1$ and

b) a survival plan effective also beyond the next 5 Gy by forming an 'ArtSun' (artificial Sun) in the Kuiper Belt and by moving Earth to this place by means of the above mentioned 'swing-by' technique.

## 2. Shadowing Earth in the next some gigayears by a parasol

The proposition to construct a parasol in Lagrange Point $L_1$ , to shadow Earth, was proposed by one of us already 20 years ago ( Seifritz, 1989). The rational for doing this instead of proposing directly the shifting of Earth's orbit to an outer one is the following: The construction of a shadowing screen in $L_1$ is simpler, better known, less complicated and less risky than the capture, the guidance and exact control of an asteroid with ~100 km diameter to pass Earth in a distance of about 10 000 km.

Our scheme looks as follows. In the next ~5 Gy only two things happen in our planetary system: 1) Sun's radius increases by about a third which is not very important but 2) Sun's luminosity increases by a factor of 1.84 which will be very awkward because an unshielded Earth would become warmer: How much?

If we write down the energy balance of Earth at present

$$(1-A_o)S_o/4 = \varepsilon_o\sigma T_o^4 \quad [\mathrm{W/cm^2}] \qquad (3)$$

And in (10 - 4.58) Gy = 5.42 Gy (see Table 1) later

$$(1-A_1)S_1/4 = \varepsilon_1\sigma T_1^4 \quad [\mathrm{W/cm^2}], \qquad (4)$$

where the LHS is essentially the visible influx of solar energy and the RHS is the emitted infrared flux of energy we can see that the relation

$$[(1-A_1)/(1-A_o)] \,.\, (S_1/S_o) = (\varepsilon_1/\varepsilon_o) \,.\, (T_1/T_o)^4 \quad [-] \qquad (5)$$

holds with A, S, $\varepsilon$, $\sigma$ and T being the corresponding albedo, solar constant, emissivity factor, the Stefan-Boltzmann constant and effective surface temperature of Earth, respectively. Since the distance Sun-Earth, a, remains unchanged, $S_1/S_o = L_1/L_o = 1.84$ according to Table 1. With $A_o = 0.3$, $S_o = 0.137$ W/cm$^2$ , $\sigma = 5.669\cdot10^{-12}$ W/(cm$^2$.K$^4$), $T_o = (273+15)$ K = 288 K, we obtain in Eq.(3) $\varepsilon_o = 0.61$ representing (as a lumped factor ) the influence of the atmosphere to fulfill Eq.(3) for the present effective surface temperature of Earth being $T_o = 15$ °C.

Since $T_1 > T_o$ we can now only speculate how large $A_1$ and $\varepsilon_1$ will be in the future 'terrestrial moist greenhouse' environment in 5 Gy from now. If we let them unchanged, in zero's approximation, we obtain the simple relation

$$(T_1/T_o)^4 = L_1/L_o = 1.84 \quad (6)$$

meaning that $T_1 = 1.165\ T_0$ or the effective surface temperature of Earth will be around 63 °C or (63-15) °C = 48 °C warmer than today.

The oceans will not yet boil at 63 °C but taking into account the seasonal swing especially the lakes will resemble a hot bath-water, aside from the fact that the atmosphere will contain much more humidity and clouds with stronger rain and storms, etc. Earth will be inhabitable.

The answer is to shield Earth from solar radiation in such a manner that we maintain the present solar constant $S_o$. The best place to do so is the inner Lagrange Point $L_1$ between Sun and Earth (Seifritz, 1989). Where the sidereal period is exactly 1 year, the same as our Earth has. In $L_1$ about $1.5\cdot10^6$ km = 0.1 AU from Earth on the connecting line to Sun the attractive forces onto a body from Sun and Earth and its centrifugal force compensate. There are other Lagrange Points but they are not of interest here).

In Fig 1 we visualized our proposal. In order to shield the cross-section of Earth homogeneously we have chosen the size of the partial transparent parasol in such a way that Earth, from the North Pole to its South Pole, lies in its deepest shadow, i.e. in its umbra.

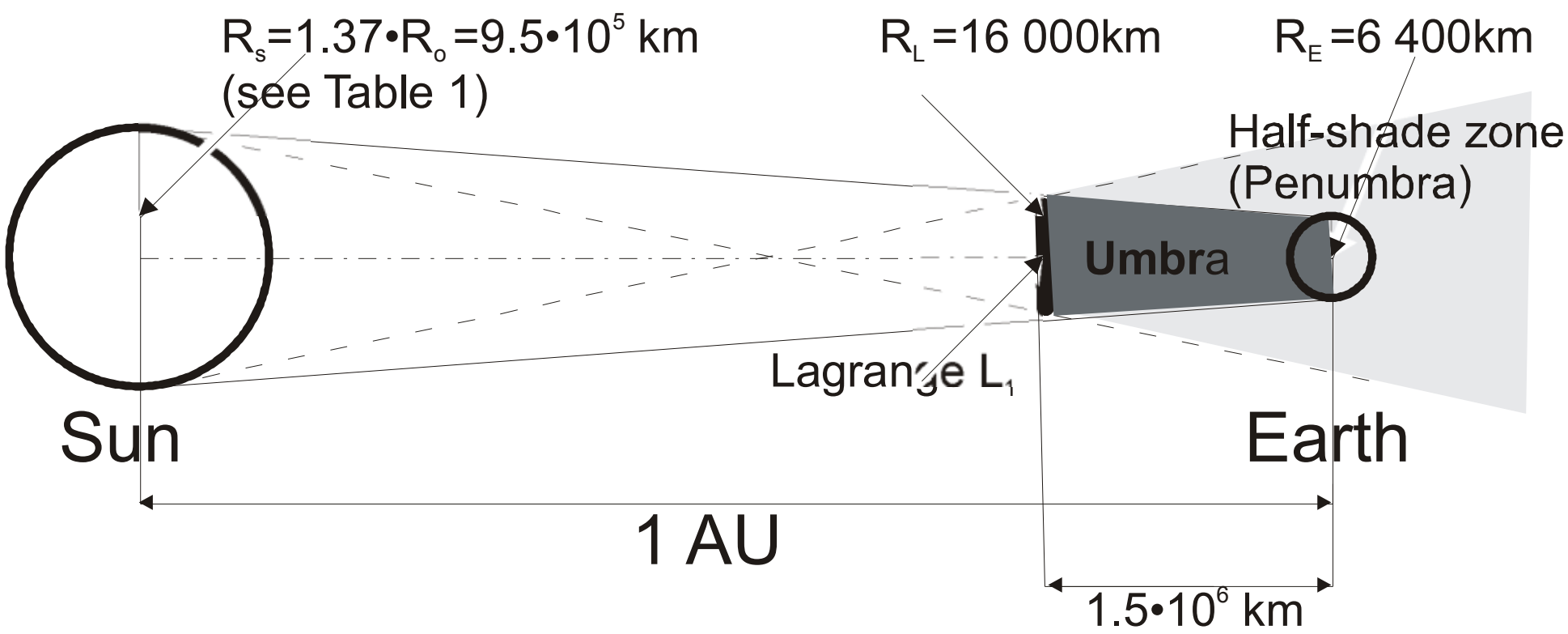

Fig.1. Partially transparent 'parasol' (opacity 0.46) in $L_1$ possessing here in this limiting case a radius of R ~16 000 km, 2.5 times Earth's radius $R_E$. Here, $R_L$ is chosen in such a way that the total cross section of Earth lies in the umbra and the half-shade zone lies completely outside Earth.

To guarantee the present solar constant, $S_0$, the transparency factor, $T_r$, of the parasol must obey

$$(L_1/L_o)\cdot T_r \equiv 1 \quad (7)$$

With Sun's luminosity ratio in about 5 Gy being 1.84 we obtain

$$T_r = 0.54_3 \qquad (8)$$

or the opacity, O, of the parasol, $O = 1 - T_r$, is 46% meaning that only 46% of its area consists really of opaque shield material. Thus, we propose on the one hand to make a complete and permanent solar eclipse, but on other hand the area of our artificial barrier in $L_1$ is covered only by 46% with an opaque material. This kind of eclipse maintains the present solar constant on Earth in 5 Gy when Sun's luminosity will be increased by a factor of 1.84. The radius $R_L$ of the perforated parasol in $L_1$ turns out to be $R_L \cong 16\,000$ km which seems to be very large but the arrangement shown in Fig.1 possesses the possibility to guarantee an optimized climate everywhere on Earth too, by adjusting its local permeability accordingly. Even in the case we let the parasol rotate (due to stability reasons) a radial adjustment of the permeability of the parasol would allow to control the solar influx on Earth from the equator to the poles adequately. Hence, not only the present global specific influx of solar energy can be maintained over the whole cross section of Earth but it could even be matched to the regional demands of Earth, too, to improve the local climate, e.g. to cool more the deserts and less the poles. To stabilize the ocean levels, for instance, it is thinkable to cool the South Pole and the Greenland area less to keep the ice on shore, etc.

If we assume that the covered area is a 1 mm thick foil the volume of the parasol is $\pi R_L^2 \cdot 1\ \mathrm{mm} \cdot 0.46 = 370\ \mathrm{km}^3$ corresponding to a spherical diameter of about 9 km. Since we have plenty of time to start with the creation of the parasol we will certainly dispose on a point of support on Moon where gravitational acceleration is only 1.62 $m^2/s$ and where the escape velocity is 2.37 km/s. Therefore, we can use the material from Moon, working it up there and shooting it to $L_1$. Probably the best suited method to accomplish this task are electromagnetic launchers in future which have been already studied over the years (Angel, 2006).

Unfortunately, the Lagrange Point $L_1$, is not an inherently stable equilibrium of microgravity (saddle point) meaning that any small deviation of a body from $L_1$ will not drive it back. Therefore, the construction has to be stabilized continuously. A dust-cloud, for instance, of a smashed asteroid would dissipate away in short time. Therefore, the second of us has given up this simple idea.

Depending on the reflectivity, R, of the surface of the parasol facing Sun, the momentum transfer and the force from the solar photons hitting the parasol will be different. Therefore, we have to take into account and additional force pressing the parasol towards Earth. M. Mautner (Mautner, 1993) and recently R. Angel (Angel, 2006) have analyzed this effect in details. The consequence is that the parasol has to be positioned a bit nearer to Sun depending on R. In any case the sidereal period of the parasol persists in one year. Such an orbit is not subject anymore to Kepler's law because of the additional consideration of a radiation pressure which cannot be neglected.

The arrangement discussed so far and shown in Fig.1, in which Earth is fully hidden in the umbra of the parasol, should be considered as a maximum solution of the problem. If we reduce $R_L$ , the umbra covers only a part of Earth's cross-section and the outer regions lie in the half-shade zone; reducing $R_L$ further Earth will lie completely in the latter. Fig.1, therefore, may be not the last word concerning the size of the parasol: a smaller but completely opaque one could probably also do the job but the possibility to adjust regionally the solar influx on Earth disappears steadily the smaller $R_L$ is. A more detailed calculation has to find the optimum where the parasol will still meet the requirements thereby minimizing its mass.

Clearly, further improvements concerning the construction of a parasol may be possible: Up to now, we have only spoken of absorbing or reflecting materials which cast a shadow on Earth. However as R. Angel (Angel, 2006) pointed out also a refractive screen deflecting the sunlight at the side of Earth rather than to absorb or reflect it, was possible. He proposed for this purpose a low reflectivity silicon nitride ceramic foil possessing broad-band anti-reflection coatings with different refractive indices thus reducing substantially the radiation pressure. Also his screen consists of a cloud of many spacecraft autonomously stabilized by modulating solar radiation pressure. These meter-sized 'flyers' would be assembled completely before launch avoiding any need for construction or unfolding in space. Each 'flyer' would weigh only a gram (Angel, 2006). Furthermore for the transport from Earth's orbit to $L_1$ Angel proposes a space-proven ion-propulsion system which would add only a small additional mass for the vehicle.

These and other proposals have to be taken into account when the shielding plan in $L_1$ will be tackled. Still a problem is the effective lifetime of any kind of parasol. The presently discussed 50 years for actual spacecraft vehicles is of course too low because we will need the parasol in the order of magnitude of one Gy. Here, new innovative ideas are especially necessary.

Since the next ~2.5 Gy Sun's luminosity increases only by 26% (see Table 1) we have plenty of time for preparation at any rate. An exercise would be to build a smaller parasol in the frame work of the present $CO_2$-greenhouse debate to compensate a temperature increase of a few degrees of centigrades on Earth (Seifritz, 1989).

In summary, it can be said that the venture discussed here and compared with Korycansky's move of Earth by successive 'swing-by' manoeuvres, is nearer at hand and less risky to survive the next 5 Gy. However, both methods will fail after this time span and life on Earth will end definitely. At least, however, we can stretch considerably the time for life on Earth with this method.

## 3. The time beyond the next 5 Gy. Planetary Engineering.

As we have already pointed out we see no realistic possibilities to survive also the last 2.5 Gy of the pulsating Red Giant phase – neither by any shielding method nor by shifting Earth's orbit.

In the following we dare a solution which, at least, is not forbidden concerning the cornucopia of Cosmos. If we want to rescue mankind as a whole together with its accustomed habitat Earth we have to swallow the bitter pill and to move Earth outside the sphere of influence of the Red Giant, i.e. into the Kuiper Belt.

One of us (Taube, 1982, Taube, 2008) has outlined the concept of the 'Rescue of Earth' . It must fulfill the following conditions.

To survive the threatening expansion of our Sun (see Table 1) during the later stages of the Red Giant we request that the whole mankind together with all of its biosphere, the whole planet and even the Moon, should be rescued and not only a 'cream' of the inhabitants of Earth, compelling the following steps.

1) Construct a artificial source of light in the Kuiper Belt illuminating the removed Earth with a 'Sun-like-light' possessing an intensity of around 1.3 kW/m$^2$ on the surface of Earth. This means a creation of an artificial long-lived star called here 'ArtSun'.

2) During the long time span of the journey from Earth's present position (1 AU) to the Kuiper Belt (~50 AU) Earth must be illuminated by the same power emitted by an artificial light source. This light source must be coupled to Earth during the whole shifting period up to the Kuiper Belt (tens of megayears)

3) The 'swing-by' technique of Korycansky should be adapted to this 'shifting' in a such way that life on Earth can endure.

### 3.1 The creation of an 'ArtSun'

The creation of the 'ArtSun' in the Kuiper Belt before the later Red Giant stages of our Sun is based on the following considerations: Within the radius of 20 light-years from our solar system there exist some thousands of planetary systems - most of them possessing Jupiter-like planets. These ‘gas-giants’ have masses being around ten times larger than the mass of Jupiter or about 1% of the mass of Sun. The chemical composition of these ‘gas giants’ corresponds to that of Jupiter ~90 atom-percent of hydrogen, almost 10 atom percent of helium and some other elements.

Some hundreds of such ‘gas giants’ will be transported to the Kuiper Belt by means of the ‘swing-by’ technique and fused together to form an ‘ArtSun’ which will ignite when its mass passes over a certain value. Unmanned spacecraft under fully autonomous control will explore those planetary systems and will find the corresponding asteroids for the ‘swing-by’ technique to accelerate the suited ‘gas giants’ out of their planetary systems. DD-fusion will be the source of energy for all these enterprises whereby deuterium will be separated out from the atmosphere of the ‘gas giant’. Although we do not know how to ignite a DD-explosive reaction for a Dyson-like space ship without the help of fissionable material we proceed on the assumption that we will have found a method in the far future.

There is the possibility to fuse together in the Kuiper Belt only 20 imported ‘gas giants’ to create a luminous star of the M-dwarf class or with about 100 ‘gas giants’ to create a G5 star, similar to Sun. For instance, the Red Dwarf Gliese 581 possesses a mass of only 0.31 $M_o$ and has a surface temperature of ~3480 K with a life time of several hundred Gy and a luminosity of 0.013 $L_o$.

Under the assumption we let rotate Earth around such a Red Dwarf illuminated with the same ‘solar constant’ as today, we find a sidereal period for Earth being only 6.91% of a year, i.e. only 25.2 days but the not yet answered question is whether the photosynthesis will work satisfactorily under the red light.

The advantage, however, if the ‘ArtSun’ will a M-dwarf is that its lifetime will be very much longer than that of our Sun, without the danger of transformation into a Red Giant. If the ‘ArtSun’ will be larger in mass, reaching about $10^{30}$ kg as our Sun, the lifetime will be limited to around 10 Gy but – on other hand – its spectrum will be similar to our solar light and there will be no likelihood for a functional disturbance of the photosynthesis.

It should be mentioned that Nature does not know the fusion of large hydrogen containing planets to form a new star. This will be a new ‘experiment’ of mankind in the envisioned planetary engineering era.

## 3.2 Illuminating Earth during its journey from 1 AU to 50 AU

During the long time of its journey from its present position (1 AU) to the Kuiper Belt (50 AU) Earth must be illuminated by the same light intensity as today from a light source coupled to Earth. This is because the distance of Earth to the Sun increases and, therefore, the natural solar constant decreases inversely proportional to the square of the distance. To maintain terrestrial life as it is, an artificial source of light must be constructed around the Earth on an orbit of 350 000 km away in form of a ring of steel and other materials.

The light source should have a similar spectrum as the Sun has and its influx on Earth should be around 175 PW. Imagine, for instance light-emitting diodes with an efficiency of around 50% and an electrical generator possessing the same thermal efficiency, the thermal power to be produced must be in the realm of 700 $PW_{th}$. Thereby, the waste heat must be irradiated into space.

Only nuclear fuels are a possibility to fuel such a giant annular power station. Again on the basis of the DD-fusion we would need about 1.2 tons of deuterium/sec or about 100 000 tons D/day. Clearly this amounts of deuterium cannot be produced on Earth. However, one can get the deuterium from the atmosphere of Jupiter. At that time rockets can be imagined to be large enough to do the work. Such a rocket will have a diameter of 25 m, a height of 140 m and a volume of 70 000 $m^3$. Since the density of liquid deuterium is rather small, 0.07 $t/m^3$, the total mass of deuterium carried on a rocket is 5 000 tons. Hence, 20 rockets/day (roughly one rocket per hour) are necessary to serve the annular DD-power station on Moon's orbit.

Assuming a 20 My voyage of Earth a total amount of $7.3 \cdot 10^{17}$ kg D would be necessary being solely a fraction of $2.4 \cdot 10^{-5}$ of the total inventory of deuterium on Jupiter (Saturn is also a source of deuterium).

As already mentioned only 25% of the primary energy can be transformed into light, the rest of 525 $PW_{th}$ must be removed as infrared waste radiation into space. Assuming a mean temperature of the cooling surface of 75°C ~350 K its specific irradiating power in according to Stefan-Boltzmann is ~850 $W/m^2$ resulting in a dramatically large total cooling area of about 600 million $km^2$ – 20% more than the surface area of Earth.

There are of course, a lot of further questions if we go into the details, the lifetime of the illuminating ring, the energy needed, the cost with respect to the gross domestic product (GDP) etc. But we will not continue on this line. Other solutions may be possible and should be evaluated further.

In Table 2 Korycansky's 'swing-by' technique to move Earth to about half-way to Mars and the 'Rescue of Earth' by shifting Earth to the Kuiper Belt are summarized.

| Table 2 Comparison of Korycansky's move of Earth to 1.5 AU and 'Rescue of Earth' to 50 AU | | | |
|---|---|---|---|
| **Parameter** | **Korycansky's 'swing by'** | **'Rescue of the Earth'** | **Ratio** |
| **Energy source** | Transfer of orbital energy from Jupiter to Earth using *swing-by*, and thereby enlarge the radius of Earth's orbit | | |
| **Reason for the operation** | Decrease Earth's warming over the next 5 gigayears | Rescue from the Red Giant beyond the next 5 gigayears | |
| **Earth before manoeuvre** | Earth-Sun distance: 1 Astronomical Unit = 150 million km | | |
| **Earth after manoeuvre** | 1.5 AU: halfway to Mars | 50 AU: edge of Kuiper Belt | |
| **Present orbital energy of the Earth** | $2.7 \cdot 10^{33}$ joules | | |
| **Energy needed to move Earth** | $8.7 \cdot 10^{32}$ joules to 1.5 AU | $2.6 \cdot 10^{33}$ joules to 50 AU | |
| **Energy boost to the Earth** | $2.4 \cdot 10^{8}$ J/kg | 10 to100 times greater? | |
| **Duration of entire operation** | 6 gigayears | 10 megayears (approximately) | |
| **Moon** | Will be lost! | Moon must be kept in the direct neighbourhood of Earth, as stabilizing factor | |
| **Venus, Mercury** | Destabilized in a short time | | |
| **Object encounter, size, km** | 100 | Up to 500 | 5 |
| **Object encounter, mass, kg** | $10^{19}$ | $10^{21}$ | $10^{2}$ |
| **Distance of encounter** | $10^{4}$ km from Earth | | 1 |
| **Velocity of encounter** | Several tens of km/s | | 1 |
| **Number of encounters** | $10^{6}$ | $10^{8}$ (arbitrary) | $10^{2}$ |
| **Encounter sequence** | Every 6000 years | Every 6 years (arbitrary) | $10^{3}$ |
| **Object's nature** | Orbit is highly eccentric | | |
| **Transfer energy of encounter** | $10^{8}$ J/kg | Not known | |
| **Energy transfer per encounter** | $10^{27}$ joules | $10^{30}$ J | $10^{3}$ |
| **Ratio: Rescue Action/Korycansky** | 1 | 100 to 1000 (approx) | |
| **Total $10^{6}$ encounters** | $10^{33}$ J | More than 1000 times | $10^{3}$ |
| **Total $10^{8}$ encounters** | | $(10^{33}$ J$) \bullet (10^{3}) = 10^{36}$ J | |
| **Power to control the object** | Deuterium-tritium; with the tritium bred from lithium | Deuterium-deuterium, 'difficult', but not forbidden | |
| **Source of propulsion** | Processing plant (lithium-tritium) and power station | Power station (deuterium-deuterium) | |
| **Isotope separation plant** | On the object | On Jupiter | |

The 'Illuminating Ring' construction should be considered as a solution for an artificial light source when moving Earth to the Kuiper Belt where it will be 'married' with an 'ArtSun'.

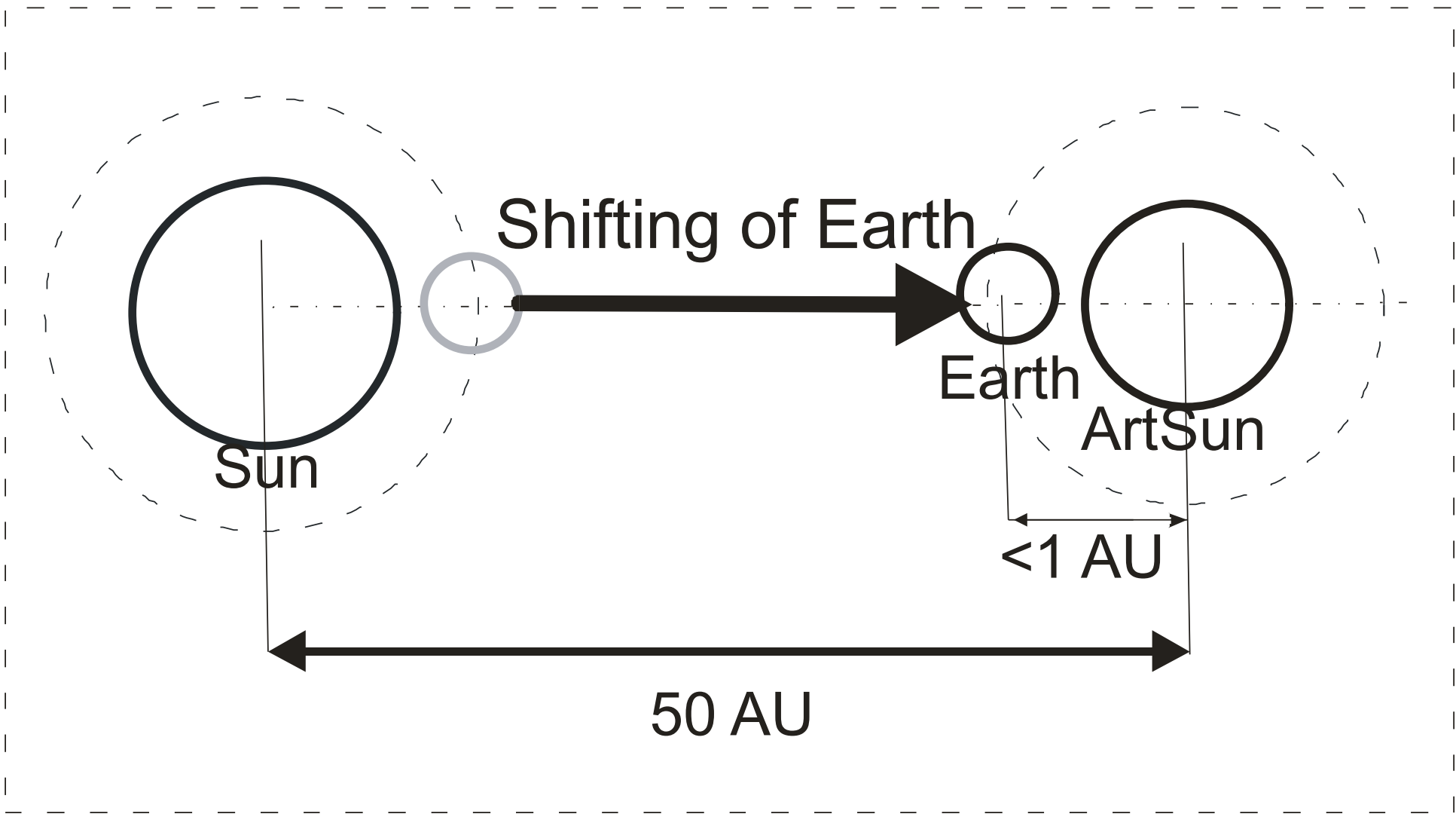

Fig 2. Earth shifted to the Kuiper Belt orbiting around ArtSun

## Conclusions

We have presented a rescue-plan for mankind to survive the Red Giant catastrophe which is subdivided into two parts:

If we content ourselves to survive the next 5 Gy only, we propose to shield Earth by a parasol rather than to move Earth to an outer orbit by Korycansky's 'swing-by' technique.

To survive the time beyond the next 5 Gy, however, we have to shift Earth into the Kuiper Belt where it will be 'married' with an 'ArtSun' formed by gaseous Jupiter-like planets 'imported' from other planetary systems within a radius of about 20 light-years. During this journey of about 10 megayears Earth must be illuminated by an artificial light source. The overwhelming energy contribution for such a 'rescue plan' stems from gravitational energy transferred by the 'swing-by' technique; for the auxiliary energy we assume that DD-fusion energy will successfully be developed in future.

It is clear, however, that there are a lot of open questions but we believe that we have disclosed a thinkable strategy how to survive the Red Giant catastrophe of our Sun without violating known laws of Nature. To our best knowledge this is the first time that somebody speculated about such a kind of Planetary Engineering.